\documentclass[prl,twocolumn,floatfix,showpacs,showkeys]{revtex4}
\usepackage{bm}
\usepackage{times}
\usepackage{amsmath}
\usepackage{amsfonts,amssymb}
\DeclareMathOperator{\grad}{grad}
\DeclareMathOperator{\dive}{div}

\usepackage{verbatim}
\usepackage{graphicx}
\begin{document}
\title{Diffusion induced mode switching in avalanching systems} 
\author{Michal Bregman}
\affiliation{Departments of  Physics,
Ben-Gurion University of the Negev,  84105 Beer-Sheva, Israel}

\begin{abstract}
Most avalanching systems in nature should involve diffusive processes as well which can change the behavior of such systems and should be taken into account. We examine the effects of diffusion on the model  
 of a dissipative bi-directional burning model. It is shown that the system behavior progressively changes with the increase of diffusion. Avalanches of small sizes become suppressed and large quasiperiodic bursts develop. In contrast with sandpiles these bursts are not edge triggered and the transition is gradual.  
\end{abstract}
\maketitle

In the last fifteen years it has been suggested that many complex physical systems behave like avalanching systems and follow the Self Organized Criticality (SOC) paradigm \cite{1988PhRvA..38..364B}. This concept proposes a new statistical mechanics for open many-body systems  with non linear dynamics, implying that such systems spend  most of the time near the critical state.
The SOC state if often characterized as the dynamical state of an avalanching system in the limit of weak driving where a number of statistical parameters are characterized by power law distributions. SOC has been suggested to occur in a large number of very different physical 
\cite{1991ApJ...380L..89L,1993PhRvB..48.4208W,1995PhRvL..74.1206F,1996PhRvE..53..414P,1997PhRvE..55.1739D,2000PhRvL..84.1192P,2001SoPh..203..321C,2004JGRA..10902218K} and non-physical \cite{1990PhRvA..41.1867M,2004PhyA..338...84L} systems and phenomena.
Among other systems where avalanching behavior is established or suspected, are a number of space  plasma systems and phenomena like solar flares and current sheet. In these systems avalanches are related to the localized magnetic reconnection. There is growing understanding that avalanches  may be a common feature or a rather wide class of systems. Most of the studies of such systems concentrate on the avalanche statistics and possible transition to SOC, without invoking additional physical mechanisms of transport. However,
there is a strong basis to believe that system dynamics can be  changed in the presence of physical mechanisms like diffusion. Most of the models that were suggested until now hardly dealt with the influence of diffusion on the dynamics of the avalanche. Yet observations show that virtually all avalanching systems in nature involve processes of diffusion, weak or strong. Such observations lead us to believe that diffusion is fundamental process in any study of dynamics of these systems.   
The first attempt to add a diffusion  flux in avalanching system was suggested in Ref.~\onlinecite{2002PhRvL..88t4304N}. The study was based on the one-directional discrete sandpile model. It was shown that the system breaks into two spatially distinct regions: diffusion dominated and avalanche dominated. For sufficiently strong diffusion the system switched into the regime of quasiperiodic large bursts, which were claimed to be edge-triggerred.  Preferential direction governs the dynamics of  the sandpile and makes the conclusions hardly applicable to other, isotropic, systems.   In the present paper we study the influence of diffusion on the avalanches in a continuous bi-directional model. We show that diffusion 
does not break the system spatially into avalanche dominated and diffusion dominated parts. However, competition between diffusion and driving can force the system to switch from the regime of usual avalanching dynamics, where  avalanches of various sizes are present,  to the  regime of large bursts with almost no small avalanches. The transition is not discontinuous but occurs gradually when avalanches of smaller sizes are suppressed as the diffusion becomes stronger.  
 
We consider the one-dimensional bi-directional   
burning model \cite{Gedalin2005a,Gedalin2005b}, which was proposed as a more accurate (albeit phenomenological) description of locally reconnecting  plasma systems.  The model is formulated in terms of energy transfer but actually was aimed to represent current carrying particles in the current sheet. The main variable is the temperature field. When the local temperature  $T(x,t)$  exceeds the upper critical value, $T>T_c$,   burning starts locally, during which energy releases at the rate $Q=kT$. Part of this energy $a$ is isotropically transferred to the neighbors while the remaining part  $(1-a)$ is radiated out of the system. This radiated energy  flux represents direct energy losses from the system and it is what  can be measured by a remote observer. Because of the  energy losses (flux to neighbors and radiation)  the local temperature decreases  until it gets below the lower critical value, $T_l$,  and the burning ceases. Thus, the smallest single site avalanche lasts for $\tau_{min}\approx \ln(T_c/T_l)/k$.  Burning can start again only when the temperature increases beyond the upper threshold. This  hysteresis behavior is the necessary feature of avalanching systems, here it represents the fact that the threshold for instability (temperature for the burning ignition or current for the reconnection triggering) is higher than what is needed for energy release maintenance. The energy loss is compensated by the external driving, that is random addition of the energy to the system. We do not assume that the driving is  weak and therefore there is no complete time separation, so that the system does not have to be in a self-organized critical state, but is a driven avalanching system.  Here we add the diffusion which proceeds independently of the threshold. 
The dynamical equations for this continuum burning model with diffusion can be written as follows:
\begin{equation}
 \frac{\partial T}{\partial t}=Q(a-1)+\dive J+\eta,\qquad J=\grad(\frac{a}{2}Q+DT),
\label{eq:flux1}
\end{equation} 
where time and length are properly normalized, 
$D$ is the constant diffusion coefficient, and $\eta$ is random external driving.
The energy release $ Q$ nonlinearly depends on the temperature and is also history dependent (hysteresis behavior):
\begin{equation}
Q_i=kT_i[\theta(T_i-T_c) + \theta(T_c-T_i)\theta(T_i-T_0)\theta(-\dot{T}_i)],
\label{eq:flux}
\end{equation}

It has been shown \cite{Gedalin2005a,Gedalin2005b} that the dynamics of the burning model differs substantially from the sandpile dynamics. Active time duration is not power-law distributed which is interpreted as absence of self-organized criticality due to the suppression of largest avalanches. As we mentioned previously the mean energy losses are due to direct radiation and not only from the boundaries. The instability criterion is local and does not depend on the behavior of the members. One of the main differences from sandpiles is that the avalanching activity occurs homogeneously throughout the system, while in a sandpile it is always enhances toward the lower end and is nearly zero (no active sites) near the top. The last is the reason for the spatially separated diffusion dominated and avalanche dominated regions observed in Ref.~\onlinecite{2002PhRvL..88t4304N} and for the edge triggering of large events.

The  nonlinear equations for the burning model dynamics cannot be solved analytically.   In order to perform numerical simulations we reformulate the model in the discrete form of 
 one dimensional cellular automaton of $L$ sites.  At each time step amount of energy $q$ enters the system at a randomly chosen position (site number $i$, $i=1,...,L$)  with the probability $p$. Thus the average energy input into a single site is $qp$ and the total time average driving into the whole system is $qpL$. The energy release from the site $i$ at the time step $t$ due to the burning is 
 \begin{equation} \label{eq:release}
\begin{split}
 Q_i(t)&= kT(t)\left[\theta(T(t)-T_c) \right.\\
 &+\left.\theta(T_c-T(t))\theta(T(t)-T_l)\theta(Q_i(t-1)\right],
 \end{split}
 \end{equation}
 and the total energy flux into site $i$ is
 \begin{equation}\label{eq:totalflux}
\begin{split}
T_i(t+1)-T_i(t)&=-Q_i + \tfrac{1}{2}(1-a)(Q_{i-1}+Q_{i+1})\\
& + D[(T_{i+1}-T_i) + (T_{i-1}-T_i)]+\eta,
 \end{split}
\end{equation}
 where the  term with $D$ is the regular diffusion. 
 The role of the diffusion  is to smooth out inhomogeneities of the temperature, thus reducing the probability of the avalanche triggering in a single site. Simplifying the explanation, let us assume that all sites are initially at some average temperature. In the absence of diffusion, addition of energy to site randomly makes the probability of triggering an avalanche a rather inhomogeneous function of position. That is, if a site starts burning, the chances to trigger burning in its neighbor are typically not high, so that small size avalanches dominate. As the diffusion increases, diffusive redistribution of energy and smoothing out the temperature differences becomes progressively faster. When diffusion is sufficiently high, the randomly added  energy is redistributed among the neighbors as a result of this diffusion, faster than energy enters the system, so that temperature of a large number of sites increases simultaneously (but more gradually). Once one of these sites starts burning the neighbors become unstable too, so that a large avalanche is triggered. In this way, small size avalanches become suppressed.  Thus,  the transition from weak diffusion to strong one will change the behavior of the system toward  larges avalanches. The diffusion at which the transition starts may be estimated as follows. Let the average temperature of the sites be $T_{av}$.  The typical time that takes to a site to become active is
$\tau_{ac}\sim (T_c-T_{av})/qp
$. 
The typical diffusive time for the propagation of the inhomogeneity to distance $L$ is     
$
\tau_{dif}\sim L^2/D$. 
Effects of diffusion should become noticeable when 
the time that takes to one site to increase his temperature and start burning becomes comparable or less than the time for the diffusive equilibration of the temperature with its neighbor:
$
\tau_{ac}\gtrsim\tau_{dif} \rightarrow    D>L^2qp/(T_c-T_{av})$,
where $L\sim 1$. At this critical diffusion the smallest, single site, avalanches start to disappear. With the increase of diffusion the size of the smallest avalanche starts to increase too. Eventually, at sufficiently strong diffusion, a typical avalanche should cover most part of the system, causing efficient draining of the energy, so that the temperature drops down to the lower threshold value $T_l$.  The system should relax to the initial state (the one before the avalanche) during the time
$
\tau_{rel}\sim (T_{av}-T_l)/qp$. 
The system becomes dominated by large bursts. Burning starting at some site proceeds in both directions adding two sites at each step since all (see below) sites are near the threshold. A site keeps burning during the time $\tau_{min}/(1-a)$ since the flux to neighbors is balanced by the back flux from them, and the only energy loss is due to radiation. Thus, we expect that the avalanche in the $(x,t)$ representation would have approximately constant with in $t$-direction. Each large burst drains the system leaving it at the temperature $T_l$. Next burst will start when the whole system is brought to the threshold $T_c$, except a part near the edge of the length $l_{e}$, where the energy accumulated during the relaxation time, $\approx l_{e}qp\tau_{rel}$ is removed by the constant diffusive flux from the edge, $\approx DT_l\tau_{rel}$, so that $l_{e}\approx DT_l/qp$. Respectively, we expect that the largest number of active size during the avalanche would be $\approx L-2l_{e}$, where $L$ is the size of the system. These estimates have been checked with numerical simulation on a medium size system of the length $L=100$.  The system parameters were chosen as follows:  the upper critical temperature $T_c=50$, the lower critical temperature $T_l=15$, the burning rate coefficient $k=0.3$, the radiative loss fraction $1-a=0.1$, the amount heat added at each step $q=2$ with the probability $p=0.005$. The probability is chosen relatively high to ensure that the system is sufficiently active.

\begin{figure}[htb]
\centering
\includegraphics[width=0.4\textwidth]{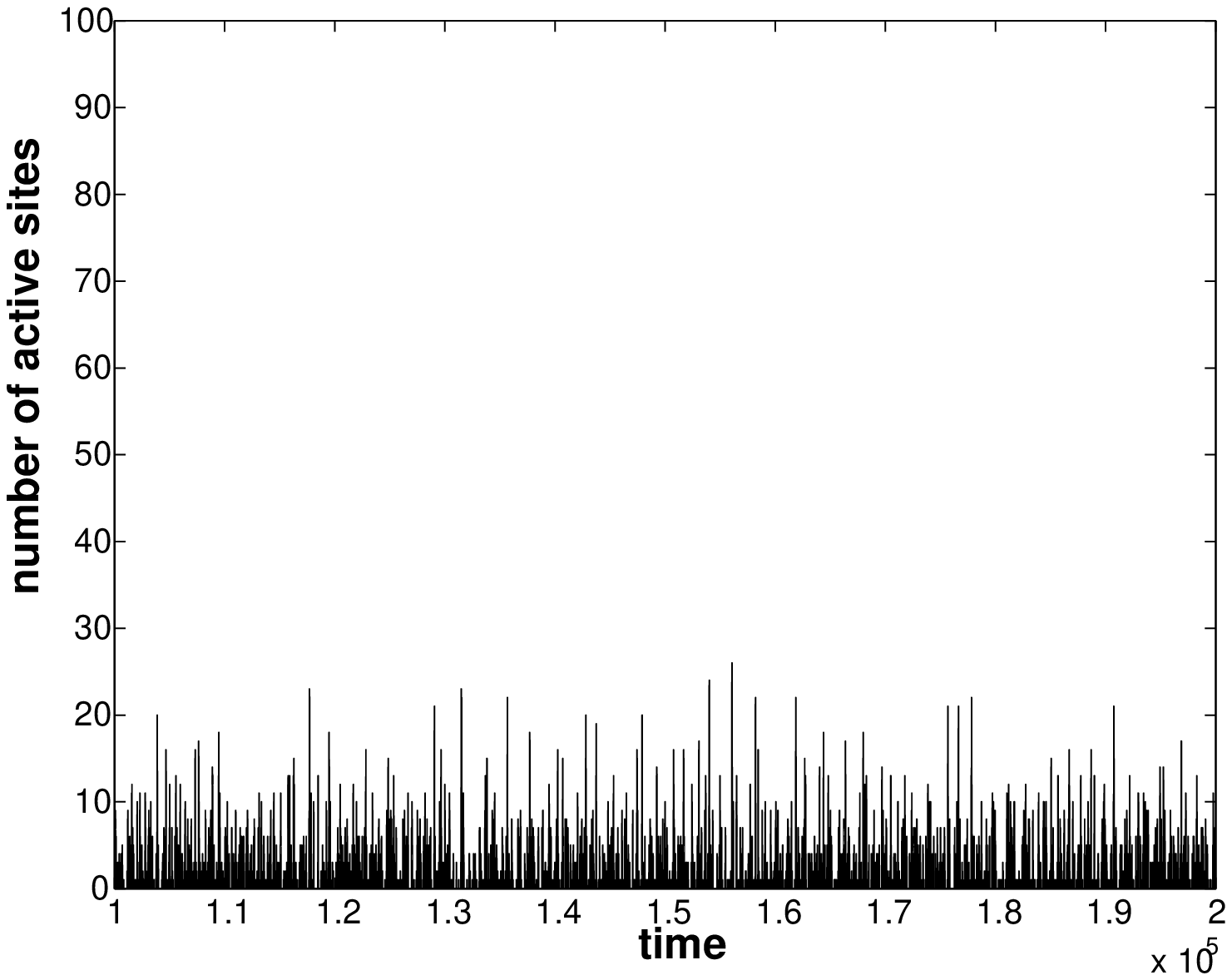}\\
\includegraphics[width=0.4\textwidth]{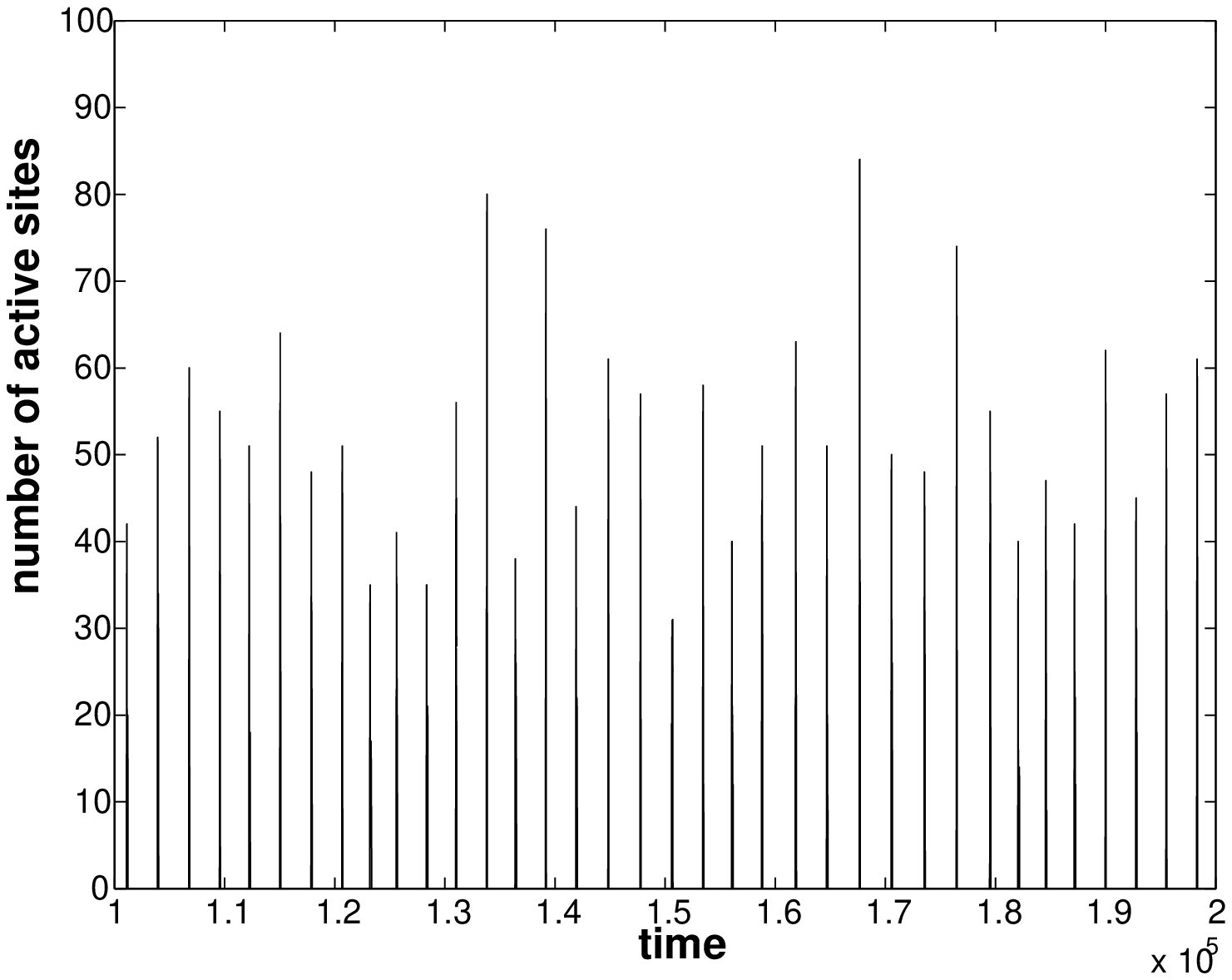}
\caption{\label{fig:avalanche_size} Avalanche size for the case without diffusion (top) and with strong diffusion (bottom).}
\end{figure}

We begin the comparison of the weak and strong diffusion with the time dependence of the avalanche size (total number of burning sites at each step). 
Figure~\ref{fig:avalanche_size} 
shows the time series for no diffusion $D=0$ (left) and strong diffusion $D=0.005$ (right).
In the case with diffusion  avalanches are less frequent than in the case without diffusion but each avalanche includes more active sites. As expected, the diffusion process causes smoothing of the temperature gradient so most of the sites have temperature around the same value and when burning begins in one site,  it eventually sweeps almost  all the system. Moreover, since the time of burning of almost whole system is the same, and the relaxation time depends only on the driving, the avalanching activity becomes quasi-periodic with the period given by $\tau_{rel}$.

\begin{figure}[htb]
\centering
\includegraphics[width=0.4\textwidth]{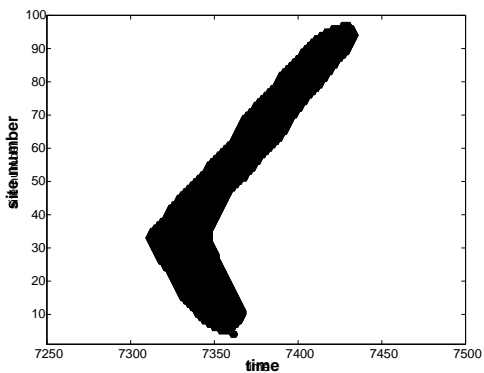}%
\caption{\label{fig:cluster} Avalanche in space and time. Width in $t$-direction and the largest number of active sites are consistent with the analytical estimates. }
\end{figure}

Figure~\ref{fig:cluster}
shows an avalanche in space and time for the strong diffusion case. The above estimates give the width of 40 time steps and the length of inactive edge regions of about 15, which agrees well with the simulation.   

\begin{figure}[htb]
\centering
\includegraphics[width=0.4\textwidth]{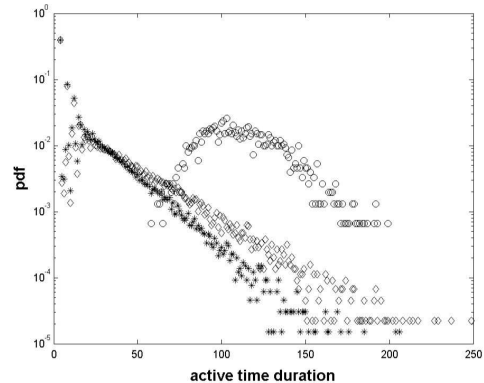}\\
\caption{\label{fig:active_time} Distribution of the active phase duration. Stars - no diffusion, diamonds - weak diffusion, and circles - strong.}
\end{figure}

One of the most accepted and reliable way to learn about the avalanching system dynamics is to study the active phase duration distribution. Figure~\ref{fig:active_time}
shows the comparison of such distribution for the system without diffusion (stars), with weak diffusion $D=0.00005$ (diamonds), and with strong diffusion $D=0.005$ (circles). It is clearly seen that weak diffusion starts to "eat up" the smallest durations. In the  strong diffusion case there are already no short periods of activity but a clear cut maximum around $t\approx 100$.    

\begin{figure}[htb]
\centering
\includegraphics[width=0.4\textwidth]{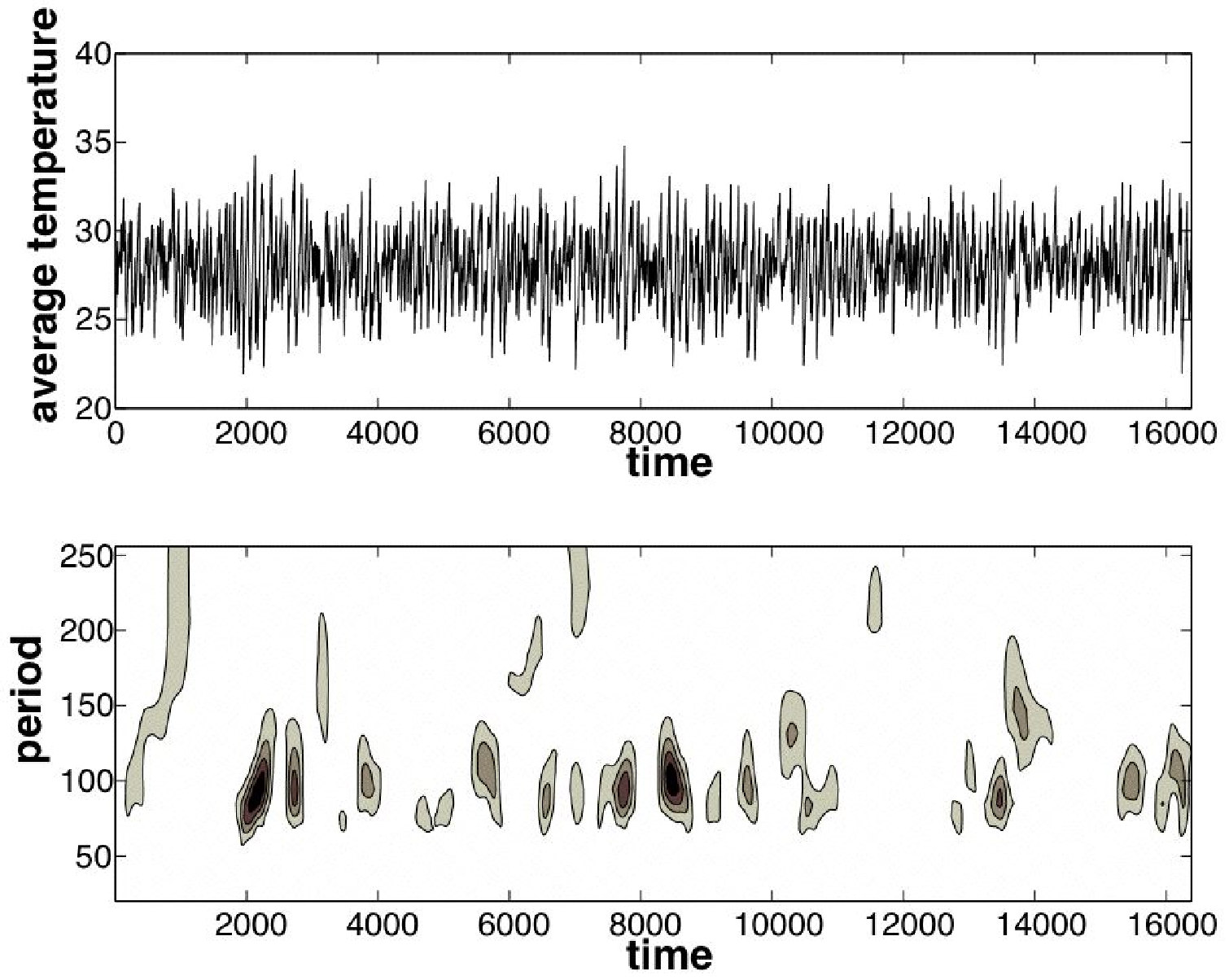}\\
\includegraphics[width=0.4\textwidth]{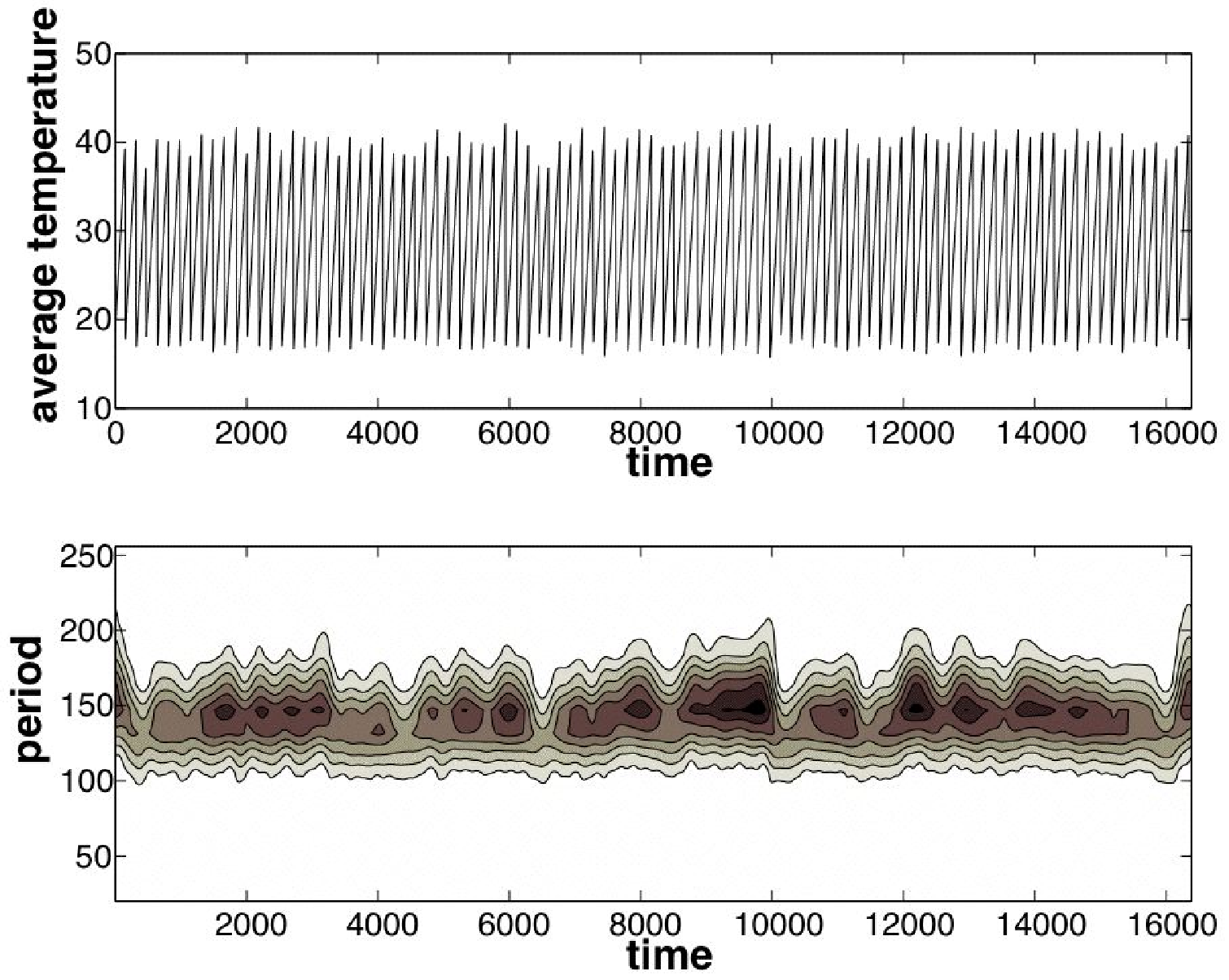}
\caption{\label{fig:average_temperature}  Average temperature as a function of time with the corresponding wavelet (Morlet) transform,  without diffusion $D=0$ (two top panels) and  with strong diffusion $D=0.005$ (two bottom panels).  Periodicity is clearly seen for the strong diffusion case.  }
\end{figure}

Yet another convenient variable for comparison is the average temperature of the system. We expect that the average temperature will fluctuate around some mean value in the case of weak diffusion. In the case of strong diffusion long periods of gradual increase should be followed by abrupt discharges. The amplitude of variations should be much larger, so that the maximum temperature should be quite close to the upper critical temperature, while the minimum temperature should be close to the lower critical temperature. Indeed, these expectations are confirmed by numerical simulations, as is seen in Figure~\ref{fig:average_temperature}. In this figure we present the average temperature with the Morlet wavelet transform\cite{Torrence1998}. 
In contrast with the Fourier transform, wavelet transform allows separation of contributions from the narrow temperature peaks themselves and from the approximate periodicity of the appearance of these peaks themselves, provided (as expected) that the width of the peaks is substantially smaller than the distance between the peaks. For our present purposes the wavelet transform presentation is enhanced to emphasize larger scales.  As we can see in the case with strong diffusion, there is clear  periodicity with the period of about $150\times16$ time steps. In the case with no diffusion we do not see such kind of behavior.

 This behavior we see also in the case of the number of active sites (Figure~\ref{fig:avalanche_size})

\begin{comment}
Finally, it was shown \cite{Gedalin2005c} that the distribution of burning clusters (connected regions of active sites) is a powerful tool in the studies of avalanching systems. 
Figure~\ref{fig:clusterdis}
\begin{figure}[htb]
\centering
\includegraphics[width=0.4\textwidth]{clusterComparison}
\caption{\label{fig:clusterdis} Cluster distribution for three cases: no diffusion (stars), weak dffusion $D=0.00005$ (diamonds), and strong diffusion $D=0.005$ (circles).}
\end{figure}
presents the comparison of the cluster distributions for three cases: no diffusion (stars), weak dffusion $D=0.00005$ (diamonds), and strong diffusion $D=0.005$ (circles). As expected, the number of small clusters greatly decreases with the increase of the diffusion coefficient. 
\end{comment}

In conclusion, in this letter we have shown that the role of the diffusion process is to smooth out the temperature distribution in the system. As a result it enhances avalanching activity  by causing large bursts at the expense of smaller ones. Strong diffusion even changes the mode of behavior of the system, where small size avalanches completely disappear and the system enters the regime of quasi-periodic bursts which span most a large part of the whole system. The maximum number of active sites is determined by the number of sites effectively drained by diffusion at the edges. The typical avalanche size is determined by the propagation of burning from the ignition point towards the edges of the active region and the duration of the burning of a single site with radiation losses only. The time between the two successive bursts depends only on the difference between the upper and lower critical temperatures and driving.      
It is important to emphasize the difference of the found quasi-periodic behavior from that described in Ref.~\onlinecite{2002PhRvL..88t4304N}. Although in both cases the basic role of the diffusion is to smooth out inhomogeneities and thus providing conditions for avalanche spreading, in the case of a sandpile the avalanching activity is related to the spatial separation of the diffusion dominated and avalanche dominated regions. Large quasi-periodic avalanches  start at the lower edge of the system and propagate up the slope to the interface between the two regions. In our case there is no spatial separation of different regions, and the spatial  distribution of activity is not triggered at the edges. The switch in the behavior of the system is not threshold like but is gradual due to the suppression of the small size avalanches for strong diffusion. Periodicity of the bursts in the strong diffusion case is related to the draining of the whole system during the large avalanche with subsequent gradual filling, and the period is determined by the external driving and separation of the upper and lower hysteresis thresholds.

\acknowledgments
The author is grateful to M. Gedalin for useful discussions and help in preparation of the manuscript.

\end{document}